\documentclass[runningheads]{llncs}
\usepackage{graphicx}
\usepackage{subcaption}

\usepackage{algorithmic}

\usepackage{xcolor}
\definecolor{LightGray}{gray}{0.9}

\usepackage{booktabs}
\usepackage{threeparttable}

\usepackage{csvsimple}
\usepackage{longtable}
\usepackage{multirow}
\usepackage{listings}

\DeclareCaptionType{listing}[Listing][List of Listings] 

\lstdefinestyle{CPP}{
	frame=single,  breaklines=true, basicstyle=\scriptsize,
	numbers=left, numberstyle=\tiny, stepnumber=1, numbersep=5pt,%
	backgroundcolor=\color{gray!10},%
}%

\usepackage[ruled,vlined,linesnumbered]{algorithm2e}

\SetCommentSty{mycommfont}

\SetAlFnt{\small}
\SetAlCapFnt{\small}
\SetAlCapNameFnt{\small}
\SetAlCapHSkip{0pt}
\IncMargin{-\parindent}

\lstdefinestyle{CPP}{
	frame=single,  breaklines=true, basicstyle=\scriptsize,
	numbers=left, numberstyle=\tiny, stepnumber=1, numbersep=5pt,%
	backgroundcolor=\color{gray!10},%
}%

\usepackage{xcolor}
\usepackage{listings}
\begin{document}
\title{Enabling Dynamic Selection of Implementation Variants in Component-Based Parallel Programming for Heterogeneous Systems}
\titlerunning{COMPAR}
%
\author{Suejb Memeti\inst{1}\orcidID{0000-0003-1608-3181}}
\authorrunning{S. Memeti}
%
\institute{Department of Computer Science (DIDA)\\
	\textit{Blekinge Institute of Technology} - 
	Karlskrona, Sweden \\
\email{suejb.memeti@bth.se}\\
}
\maketitle              
\begin{abstract}
Heterogeneous systems, consisting of CPUs and GPUs, offer the capability to address the demands of compute- and data-intensive applications. However, programming such systems is challenging, requiring knowledge of various parallel programming frameworks. This paper introduces COMPAR, a component-based parallel programming framework that enables the exposure and selection of multiple implementation variants of components at runtime. The framework leverages compiler directive-based language extensions to annotate the source code and generate the necessary glue code for the StarPU runtime system. COMPAR provides a unified view of implementation variants and allows for intelligent selection based on runtime context. Our evaluation demonstrates the effectiveness of COMPAR through benchmark applications. The proposed approach simplifies heterogeneous parallel programming and promotes code reuse while achieving optimal performance.

\end{abstract}
%
%
%
\section{Introduction}
\label{sec:introduction}

Heterogeneous parallel computing systems, comprising CPUs and GPUs, have emerged as powerful platforms capable of meeting the requirements of compute- and data-intensive applications. However, programming such systems is a challenging task that demands knowledge of various parallel programming frameworks. Multi-core resources, such as CPUs, require frameworks like Pthreads or OpenMP, while many-core resources, such as GPUs, necessitate frameworks like OpenCL or CUDA. Moreover, for many algorithms, multiple implementations exist, written using different programming frameworks or targeting different architectures.

In the era of AI advancements, tools like Github Copilot have revolutionized the development process by suggesting code snippets, functions, and algorithms directly within the editor. This capability enables developers to effortlessly create multiple implementation variants of functions, aligning with the concept of component-based programming. These tools facilitate exploration of different algorithmic approaches and code optimization for heterogeneous systems.

Despite the availability of numerous implementation variants for specific functions, the challenge lies in determining the most suitable one at a given time. Choosing the implementation variant that achieves the best performance relies on the runtime context, including input size, processing capability of available resources, and other system configuration parameters. Therefore, selecting the best implementation variant at compile-time is not feasible.

To address this challenge, this paper proposes an approach that enables developers to easily expose available implementation variants to the runtime system, which can then make informed decisions based on the given context. We introduce compiler directive-based language extensions that allow developers to annotate the source code with implementation variant options. A pre-compiler analyzes these annotations, performs syntax and semantic analysis, and generates the necessary glue code to seamlessly integrate with the StarPU runtime system \cite{augonnet2011starpu}. The StarPU runtime system takes over the decision-making process at runtime, selecting the most appropriate implementation variant based on the specific runtime context.

Several existing approaches have been proposed to expose multiple variants to runtime systems. For instance,  \cite{Usman2012} and \cite{benkner2011peppher} utilize XML-based descriptors for annotating various components, but these approaches lack seamless integration with the programming language and incur additional overhead for writing and parsing XML descriptors. Another relevant approach is the \texttt{declare variant} directive introduced in the recent version of OpenMP. Although it enables the specification of alternate implementations for base functions, it primarily supports single types of functions and lacks extensibility for diverse target architectures. Our proposed solution is inspired by the \texttt{declare variant} directive in OpenMP but extends its capabilities to support multiple types of functions and a wide range of target architectures.

The key contributions of this paper include:

\begin{itemize}
	\item A language extension that enables developers to annotate the source code and expose multiple implementation variants.
	\item A pre-compiler that performs syntax and semantic analysis and generates the necessary glue code.
	\item Utilization of the StarPU runtime system for intelligent selection of the most suitable implementation variant at runtime.
	\item An empirical evaluation of the proposed solution using various benchmark applications.
\end{itemize}

This paper is structured as follows: Section \ref{sec:compar} provides an overview of the COMPAR framework and its key components. In section \ref{sec:evaluation} we present an empirical evaluation of the proposed solution using various benchmark applications. Section \ref{sec:rw} summarizes, synthesizes, compares, and contrasts the related state-of-the-art. Finally, Section \ref{sec:conclusion_fw} concludes the paper and outlines future directions for research and development.

\section{COMPAR: Language extensions for exposing multiple component implementation variants to the runtime system}
\label{sec:compar}

This section presents the language extensions used in COMPAR to expose multiple component implementation variants to the runtime system. Furthermore, it describes the design aspects, including the syntax and usage of the COMPAR directives; and the implementation aspects, including the source-to-source compiler and the runtime system.

\subsection{Design aspects}
\label{sec:compar_design}

Benchmark studies, such as the research conducted by Memeti et al. \cite{memeti2017}, suggest that programming with OpenMP is considered more straightforward in comparison to other parallel programming models like CUDA and OpenCL. This is primarily due to the level of abstraction provided by OpenMP, which shields programmers from certain low-level details while maintaining performance that is on par with alternative models. It is important to note, however, that this abstraction may limit the level of control available for fine-tuning.

In recent versions of OpenMP \cite{openmp5.0}, the \texttt{declare variant} directive was introduced to enable the specification of alternative implementations for specific base functions. This directive, in conjunction with the \texttt{match} clause, allows for the explicit definition of the contextual conditions under which each variant should be considered. During runtime, when a function call's context aligns with that of a variant, the variant becomes a potential replacement for the base function. The selection process for determining the most suitable variant involves considering a score-based evaluation among all compatible variants.

The design of the COMPAR language draws inspiration from the OpenMP language, while aiming to expand upon the functionality offered by the \texttt{declare variant} directive to support multiple target architectures and programming models. COMPAR seeks to enhance and extend the capabilities provided by the \texttt{declare variant} directive, enabling its usage across diverse target architectures and programming paradigms.

COMPAR encompasses two primary compiler directives: \texttt{method\_declare} and \texttt{parameter}. The \texttt{method\_declare} directive is employed to annotate methods that represent implementation variants, while the \texttt{parameter} directive is utilized to define the parameters of these methods.

When using the \texttt{method\_declare} directive for the first implementation variant of an interface (i.e., function), a corresponding \texttt{parameter} directive is expected for each parameter. However, for subsequent implementation variants of the same interface, it is unnecessary to use the \texttt{parameter} directive since these variants are assumed to have the same method signature.

The syntax of the \texttt{method\_declare} directive is demonstrated in Listing \ref{lst:compar_method_declare_directive}. The directive supports several clauses, including \texttt{interface}, \texttt{target}, and \texttt{name}. The \texttt{interface} clause is used to specify the name of the interface (i.e., function) to which the variant corresponds, such as "\texttt{sort}". The \texttt{name} clause is utilized to indicate the name of the function that represents the implementation variant, for instance, "\texttt{bubble\_sort}", "\texttt{merge\_sort}", and so on. The \texttt{target} clause is employed to denote the target programming model in which the variant is written, such as \texttt{CUDA, OpenMP, Seq}, or \texttt{OpenCL}.

\begin{listing}[h]
\includegraphics[width=\linewidth]{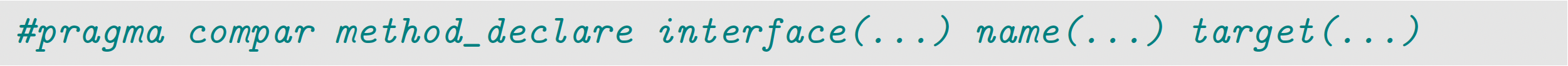}
\caption{An example of the COMPAR \texttt{method\_declare} directive.}
\label{lst:compar_method_declare_directive}
\end{listing}

The syntax of the \texttt{parameter} directive is depicted in Listing \ref{lst:compar_parameter_directive}. The directive supports various clauses, including \texttt{name, type, size,} and \texttt{access\_mode}. The \texttt{name} clause is used to specify the name of the parameter. The \texttt{type} clause indicates the type of the parameter, such as \texttt{int, float, double, char, wchar\_t,} and so on. The \texttt{size} clause indicates the size of the parameter. It is worth noting that the \texttt{size} clause can accept different numbers of parameters: one parameter for vectors, two parameters for matrices, three parameters for 3-dimensional data structures, and four parameters for 4-dimensional data structures. Lastly, the \texttt{access\_mode} clause corresponds to the access mode of the parameter, such as \texttt{read, write}, or \texttt{readwrite}.

\begin{listing}[h]
\includegraphics[width=\linewidth]{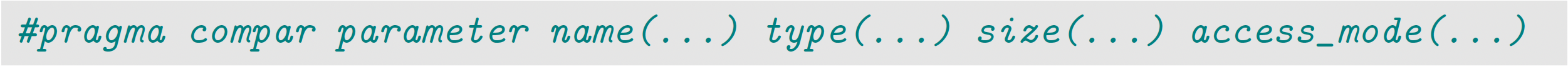}
\caption{An example of the COMPAR \texttt{parameter} directive.}
\label{lst:compar_parameter_directive}
\end{listing}

Listing \ref{lst:compar_example} presents an excerpt from an application that demonstrates the use of the COMPAR framework with the \texttt{sort} and \texttt{matrix multiplication} functions, each having two implementation variants. Specifically, for both functions there are corresponding CUDA and OpenMP implementation variants.

For the sort function, two parameters are utilized: an array of floats and a scalar integer. In the case of the matrix multiplication interface, four parameters are required: two 2-dimensional float arrays (\texttt{A} and \texttt{B}) with a size of \texttt{N $\times$ M}, as well as two scalar integers, \texttt{N} and \texttt{M}.

Within Listing \ref{lst:compar_example}, lines 23 and 24 exemplify method calls to the defined interfaces, specifically invoking the \texttt{sort} and \texttt{mmul} functions, respectively.

\begin{listing}
\includegraphics[width=\linewidth]{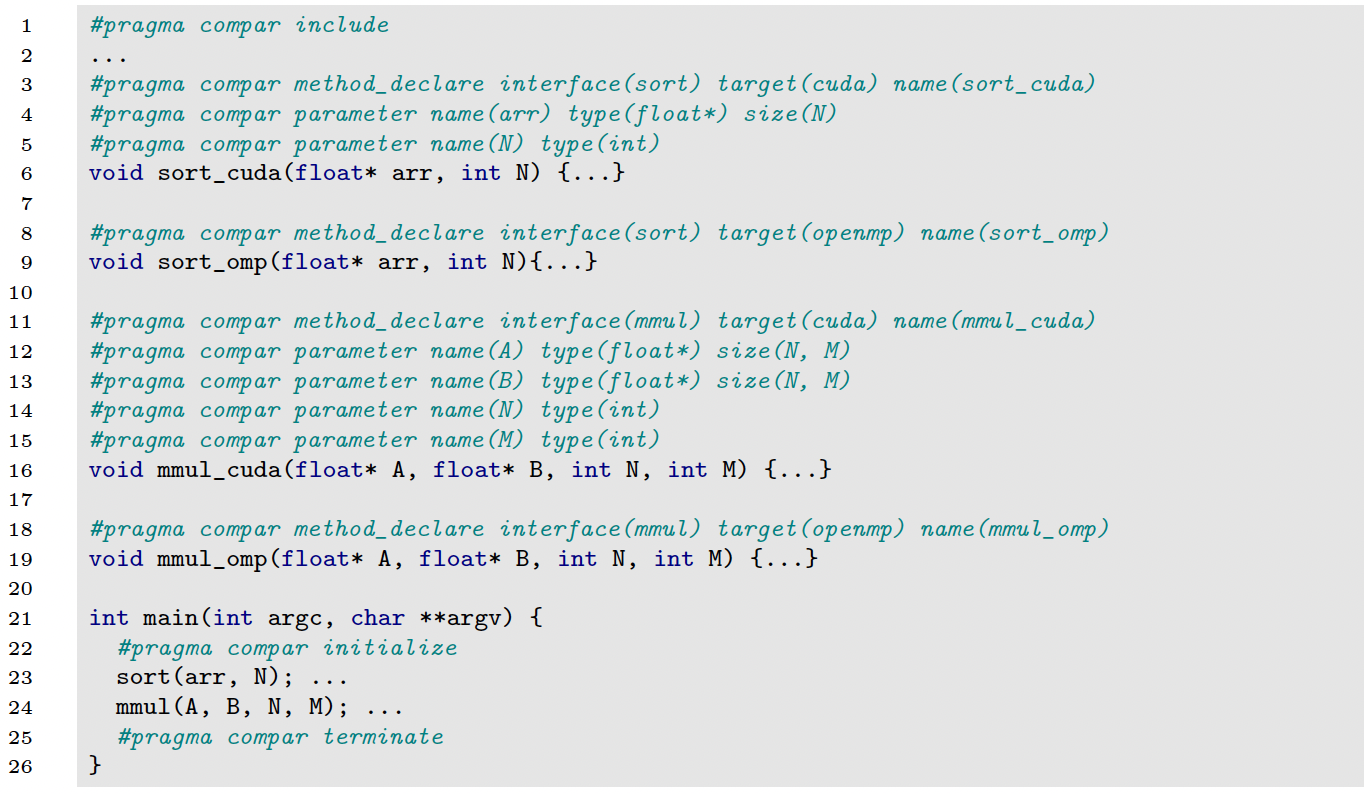}
\caption{Source code example of using COMPAR to expose multiple implementation variants of \texttt{sort} and \texttt{mmul} to the runtime system.}
\label{lst:compar_example}
\end{listing}

Lines 1, 22, and 25 in Listing \ref{lst:compar_example} demonstrate additional COMPAR directives, which are seamlessly translated into their corresponding C/C++ code. For instance, the \texttt{\#pragma compar include} directive is translated to \texttt{\#include "compar.h"}, ensuring that the necessary COMPAR functionality is included.

Similarly, the \texttt{\#pragma compar initialize} directive is translated into a method call, \texttt{compar\_init()}, which is defined within the generated \texttt{compar.h} file. Invoking this method call initializes the COMPAR framework, preparing it for utilization within the application. Likewise, the \texttt{\#pragma compar terminate} directive is transformed into a method call, \texttt{compar\_terminate()}, also defined in the \texttt{compar.h} file, ensuring the proper termination and cleanup of the COMPAR framework at the conclusion of program execution.

It is important to note that all of the COMPAR directives, if not processed by our pre-compiler, do not introduce any changes to the existing code. Thus, the original code would continue to function as intended, ensuring backward compatibility is maintained.

\subsection{Implementation Aspects}
\label{sec:compar_implementation}

This section discusses the tools and techniques employed in the implementation of the source-to-source pre-compiler for COMPAR. The pre-compiler is responsible for translating COMPAR pragma directives into the corresponding C/C++ code, including the integration of the StarPU runtime system.

\subsubsection{Source-to-Source Compiler}

The COMPAR pre-compiler encompasses various phases of a compiler, including lexical, syntax, and semantic analysis (the front-end), and intermediate representation and code generation (the back-end).

For the lexical analysis, the Flex tool (Fast Lexical Analysis Generator, formerly known as Lex) was utilized to define and write the COMPAR language specification. Since COMPAR is a pre-compiler, it only needs to analyze the parts of the program that start with \texttt{\#pragma compar}. Therefore, the language specification is straightforward.

To perform syntax analysis, the GNU Bison tool (formerly known as Yacc) was employed to define and write the syntax specification for COMPAR language extensions. The syntax analyzer ensures the correct structure and usage of COMPAR directives, validating the values and clauses provided within them. It generates an abstract syntax tree for further processing.

The semantic analysis phase verifies the semantic correctness of the COMPAR directives within their respective contexts. It checks for duplicate interface or parameter definitions and ensures the correct usage of clauses and options. While the current version of COMPAR makes certain assumptions, such as assuming the existence of variable names provided in the clauses, additional analysis steps are needed in a production environment to enforce such requirements.

The current version of COMPAR does not include compile-time optimization. However, as mentioned in the future works section (Section \ref{sec:conclusion_fw}), optimization techniques could be applied during compilation to reduce the set of implementation variants based on benchmarking results or other criteria.

Assuming no semantic errors are found, the compiler proceeds to the intermediate representation (IR) phase, where the IR is generated, capturing the necessary information for subsequent code generation. The code generator, utilizing template-based techniques, then produces the target code, in this case, the StarPU code and the glue code required to integrate the source code with the StarPU runtime system \cite{augonnet2011starpu}. It is important to note that StarPU is considered a back-end target tool, and it can be easily replaced with other runtime systems, such as StarSs \cite{ayguade2009extension}.

\subsubsection{Runtime System}

COMPAR utilizes the  StarPU runtime system \cite{augonnet2011starpu} to handle the mapping and execution of the various implementation variants on different computational resources, including CPUs and accelerators.

StarPU operates on a task-based model, where applications submit computational tasks, each with multiple potential implementations targeting different heterogeneous processing units. The StarPU runtime system handles the mapping, scheduling, and data transfers required for executing these tasks. The key components in StarPU are codelets and tasks. A codelet in StarPU corresponds to a variant implementation in COMPAR, representing different implementations of the same algorithm targeting various architectures. A task in StarPU, similar to a function interface in COMPAR, represents a set of codelets and the data required to execute them. When a task is executed, a codelet is selected based on the specific architecture and data associated with it.

The COMPAR pre-compiler generates the necessary code to define codelets, input/output data parameters, and task submission for the StarPU runtime system. 
For example, in the code excerpt in Listing \ref{lst:compar_starpu_example}, which corresponds to the example in Listing \ref{lst:compar_example}, the generated code integrates the StarPU runtime system.

\begin{listing}
\includegraphics[width=\linewidth]{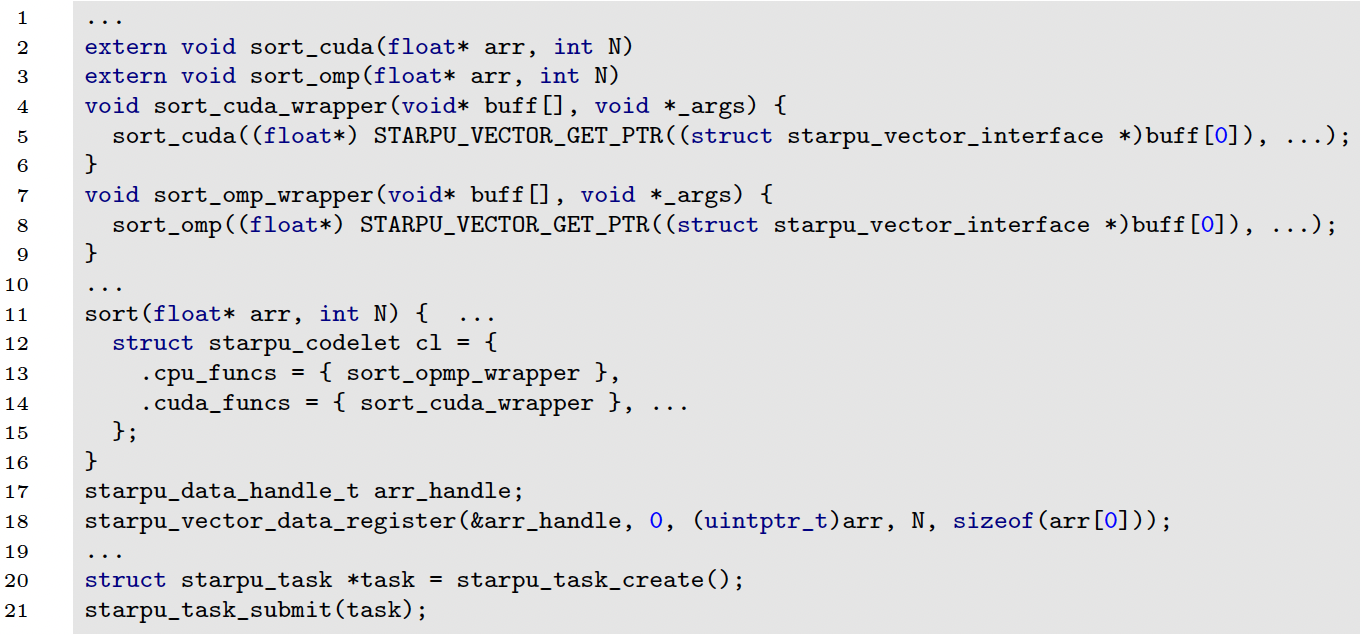}
\caption{Excerpt of the COMPAR generated code for StarPU. The generated code corresponds to the example shown in Listing \ref{lst:compar_example}}
\label{lst:compar_starpu_example}
\end{listing}
Lines 2-9 declare the external functions for sorting using CUDA and OpenMP and define their corresponding wrapper functions, which are then specified as implementation variants of the StarPU codelet inside the \texttt{sort} function (lines 11-16). Lines 17-18 show the code for registering the data handle for the input array \texttt{arr}. Lines 20-21 show the code corresponding to the creation and the submission of the task. Note that all data handle(s) associated with the parameter(s) need to be unregistered, which is not shown on the example.

It is important to note that this example showcases the generated code for the \texttt{sort} function, while the example in Listing \ref{lst:compar_example} includes another interface named \texttt{mmul} for matrix multiplication. However, the corresponding generated code for the \texttt{mmul} interface is not presented in the example. COMPAR generates separate code files, similar to the code excerpt in Listing \ref{lst:compar_starpu_example}, for each defined interface.

\section{Evaluation}
\label{sec:evaluation}

This section, first, describes the experimentation environment, which includes details related to the hardware configuration, the set of application benchmarks and the used data-sets for the corresponding applications, as well as the evaluation metrics. Afterwards, the results of the empirical evaluation are presented and discussed.

\subsection{Experimentation environment}
\label{sec:exp_environment}


Table \ref{table:sys_config} lists the properties of the heterogeneous computing system that was used for experimentation in this paper. 

\begin{table}[ht]
	\scriptsize
	\vspace{-15pt}
	\caption{Hardware system configuration.}
	\label{table:sys_config}
	\centering
	\begin{tabular}{ p{3.5cm} p{4.15cm} p{4.15cm} }
		\toprule
		& Multi-core CPU & Many-core GPU \\
		\midrule
		Processor & Intel Xeon E5-2620 v4 & NVIDIA GP102 Titan Xp \\
		\# cores and core frequency & 8 cores, 2.10 - 3.00 GHz & 3840 cores, 1.41-1.58 MHz \\
		Cache size & 20 MB Intel Smart Cache & L1: 48KB per SM; L2: 3MB  \\
		Memory size and bandwidth & 96GB, 68.3 GB/s & 12GB, 547.6 GB/s \\
		Thermal Design Power & 85 W & 250 W \\
		\bottomrule
			\vspace{-15pt}
	\end{tabular}
\end{table}


To empirically evaluate the performance and the programmability effort required to utilize the COMPAR framework, we have selected various benchmarks from the Rodinia benchmark suite, including the  hostpot, hotspot3D, lud, nw, and a matrix multiplication application which is not part of the Rodinia benchmark suite.

Table \ref{table:benchmarks} lists the benchmark applications and the different ranges of input that were used during the evaluation. Each of the configuration, i.e. hardware platform selection and input configuration is repeated for 10 times, and the average values are reported.

\begin{table}[ht]
	\scriptsize
	\begin{threeparttable}
		\caption{Benchmark applications used to evaluate COMPAR.}
		\label{table:benchmarks}
		\centering
		\begin{tabular}{ p{2.5cm}p{4.5cm}p{3cm}p{1.8cm}}
			\toprule
			Application 
			& Implementation variants 
			& Input parameters\tnote{*} 
			& Input range \\
			\midrule
			Hotspot 
			& CUDA, OMP 
			& squared grid size 
			& 64 - 8192
			\\
			Hotspot3D 
			& CUDA, OMP 
			& rows/cols 
			& 64 -  512 
			\\
			Lud 
			& CUDA, OMP 
			& squared matrix size 
			& 64 - 8192 
			\\
			Nw 
			& CUDA, OMP 
			& max. rows/cols 
			& 64 - 8192 
			\\
			Matrix multiply 
			& BLAS, OMP, CUDA, CUBLAS 
			& squared matrix size 
			& 8 - 8192
			\\
			\bottomrule
		\end{tabular}
		\begin{tablenotes}
			\scriptsize
			\item[*] Only parameters that are used to scale the application are shown here.
		\end{tablenotes}
	\end{threeparttable}
\end{table}

\subsection{Results}
\label{sec:results}

\begin{figure}
	\centering
	
	\begin{subfigure}{0.49\textwidth}
		\centering
		\includegraphics[width=\linewidth]{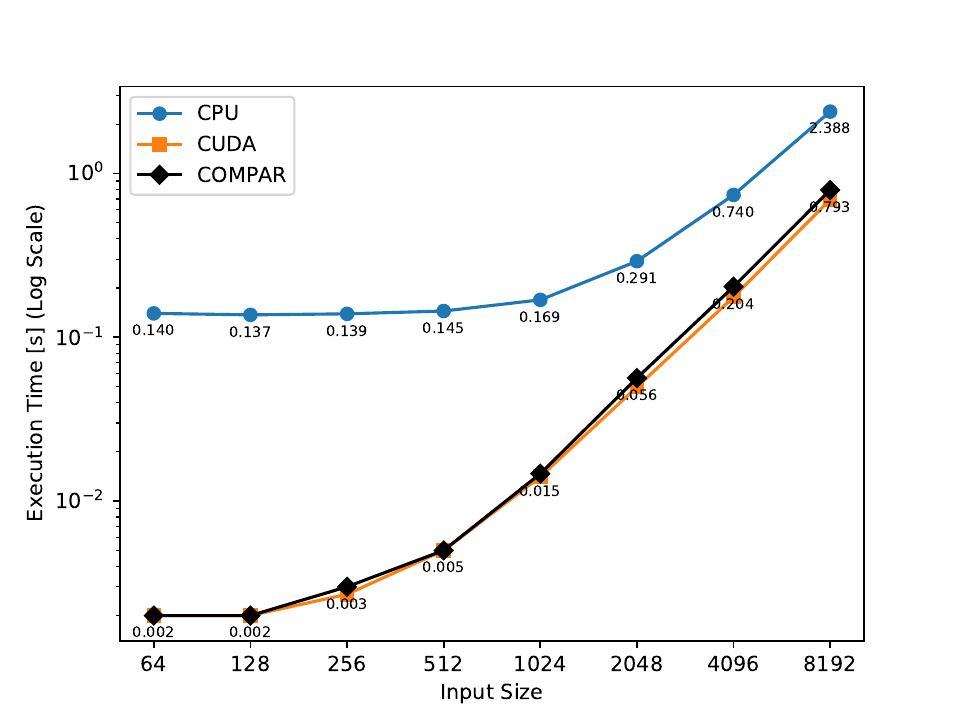}
		\caption{Hotspot}
		\label{fig:hotspot}
	\end{subfigure}
	\hfill
	\begin{subfigure}{0.49\textwidth}
		\centering
		\includegraphics[width=\linewidth]{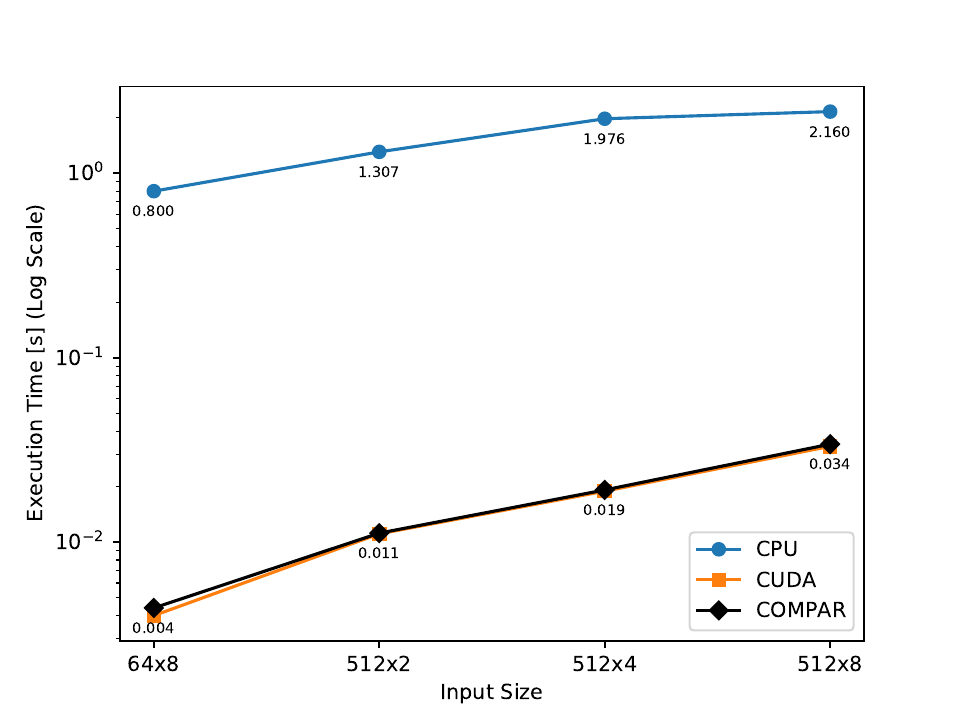}
		\caption{Hotspot3D}
		\label{fig:hotspot3d}
	\end{subfigure}
	
	\vspace{0.5cm}
	
	\begin{subfigure}{0.49\textwidth}
		\centering
		\includegraphics[width=\linewidth]{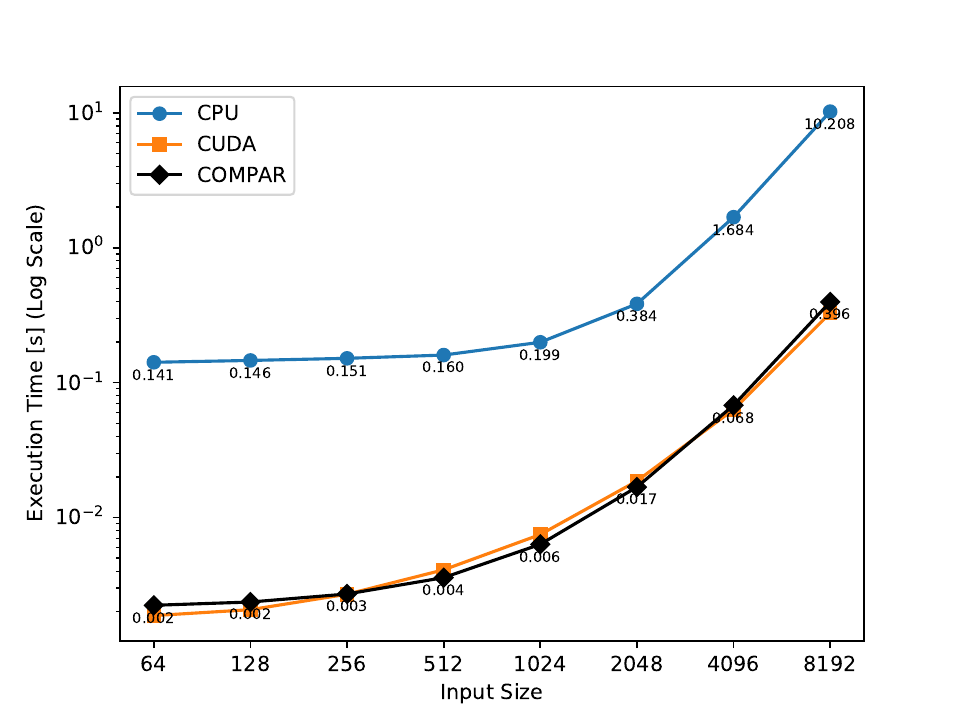}
		\caption{LUD}
		\label{fig:lud}
	\end{subfigure}
	\hfill
	\begin{subfigure}{0.49\textwidth}
		\centering
		\includegraphics[width=\linewidth]{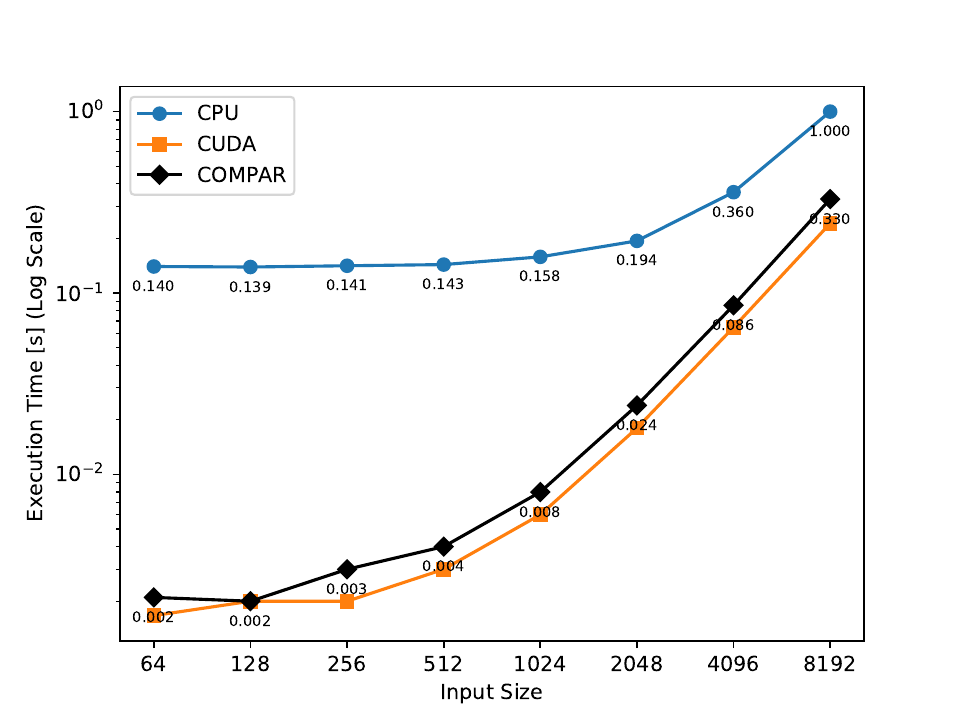}
		\caption{NW}
		\label{fig:nw}
	\end{subfigure}
	
	\vspace{0.5cm}
	
	\begin{subfigure}{0.49\textwidth}
		\centering
		\includegraphics[width=\linewidth]{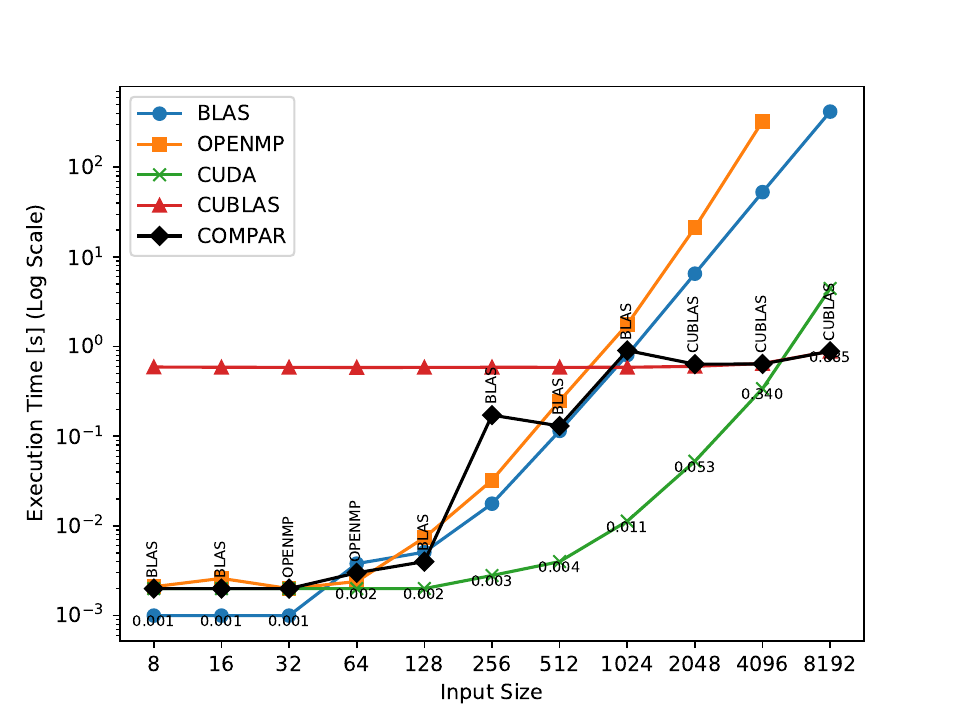}
		\caption{Matrix Multiplication}
		\label{fig:mm}
	\end{subfigure}
	\hfill
	\begin{subfigure}{0.49\textwidth}
		\centering
		\begin{tabular}{l c c c}
			\toprule
			Application & StarPU & \cite{Usman2012} & COMPAR\\ \midrule
			hotspot 	&	447		&	327	&	7 \\
			hotspot3D 	&	/		&	/	&	9 \\
			lud 	&	586		&	510	&	7 \\
			nw 	&	449		&	359	&	8 \\
			mmul 	&	229		&	140	&	14 \\
			\bottomrule
		\end{tabular}
		\caption{Comparison of the Lines of Code (LOC) authored by programmers when utilizing StarPU, the methodology proposed in \cite{Usman2012}, and COMPAR. The metrics for StarPU and the approach outlined in \cite{Usman2012} are sourced from the referenced report.}
		\label{fig:loc}
	\end{subfigure}
	\caption{Evaluation results}
	\label{fig:eval_results}
\end{figure}

The evaluation results of the COMPAR framework are presented in Figure \ref{fig:eval_results}, where a comparison is made between CPU-only and GPU-only executions. The CPU-only configuration is controlled by setting the \texttt{STARPU\_NCUDA} environment variable to 0, while the GPU-only configuration is controlled by setting the \texttt{STARPU\_NCPU} environment variable to 0.

In this specific hardware configuration, where the GPU exhibits significantly higher performance compared to the CPU, most of the benchmark applications (except for matrix multiplication) demonstrate improved performance when executed on the GPU. However, it is important to note that this observation may not hold true for other hardware configurations or applications. The matrix multiplication application serves as an example, as depicted in Figure \ref{fig:mm}. For smaller input sizes (8-128), it is not always clear which implementation variant (BLAS, OPENMP, or CUDA) performs the best. In the case of matrices with dimensions of 4096, the performance of the CUDA implementation surpasses that of the CUBLAS variant. Conversely, when the matrix dimensions are expanded to 8192, CUBLAS demonstrates superior performance. This emphasizes the non-trivial nature of such decisions, suggesting that they should be delegated to the runtime system rather than being hard-coded. 

The empirical evaluation of the COMPAR framework, as illustrated in Figures \ref{fig:hotspot}-\ref{fig:nw}, reveals that the code generated by COMPAR, when integrated with the StarPU runtime system, consistently opts for the most performance-efficient implementation variant. It is important to highlight that the marginal discrepancies in execution time between COMPAR and a CUDA-only approach can be ascribed to the stochastic variability inherent in performance experiments. Although the CUDA-only implementation frequently exhibits superior performance—likely due to the absence of overheads associated with StarPU's decision-making mechanism—there are specific instances (e.g., in the LUD application) where the COMPAR version outperforms its CUDA-only counterpart.

In relation to the matrix multiplication application, which involves multiple implementation variants, it was observed that the STARPU selection mechanism frequently chose sub-optimal options. For example, while the BLAS implementation is the optimal choice for matrices of size 32, COMPAR—guided by the STARPU runtime—opted for the OPENMP variant. Likewise, for matrix dimensions ranging from 64 to 4096, the CUDA implementation demonstrated superior performance; however, STARPU selected less efficient variants, such as OPENMP and BLAS. Given that the STARPU decision-making process relies on machine learning models, it is reasonable to hypothesize that additional training of these models could lead to more accurate and optimal variant selection.

Table \ref{fig:loc} presents findings related to programmer productivity, also referred to as programmability. It is evident that the COMPAR approach necessitates substantially less effort than both the method proposed in \cite{Usman2012} and direct usage of StarPU. It should be noted that the metrics for the aforementioned approaches are derived from the study by Dastgeer et al. \cite{Usman2012}. Additionally, it is worth mentioning that results for the \texttt{hotspot3D} application are absent, as this specific application was not evaluated in the study by Dastgeer et al. \cite{Usman2012}.

\section{Related work}
\label{sec:rw}

SYLKAN \cite{Thoman2021} is a framework that extends the SYCL programming model to utilize the Vulkan graphics API for efficient parallel computing on GPUs and other Vulkan-compatible devices. It aims to provide a Vulkan compute target platform for SYCL. While both SYLKAN and COMPAR share similar goals, they differ in focus and approach. COMPAR emphasizes component-based parallel programming for heterogeneous systems with integration of the StarPU runtime system, while SYLKAN specifically targets Vulkan as the runtime system and extends SYCL to leverage Vulkan's low-level capabilities for parallel computing.

In their work, Österlund and Löwe \cite{osterlund2018} present an approach for runtime selection of implementation variants based on varying contexts, which shares similar goals as COMPAR. While their focus lies on addressing the optimal variant selection process, our paper shifts the responsibility of selection to the runtime system. Instead, our focus is on facilitating the exposure of implementation variants to both the compiler and the runtime system.

Soudris et al. \cite{Soudris2018}, as part of the EXA2PRO project, propose a framework for exascale systems that utilizes multiple implementation variants. This idea was initially introduced in the PHEPPER project \cite{benkner2011peppher,Usman2012}. Like our work, they annotate implementation variants and rely on external runtime systems such as STARPU for selection. However, their approach requires developers to provide XML-based descriptions of the hardware platform, implementation variants, and compilation process. In contrast, our approach utilizes compiler directives for annotation, resembling the OpenMP parallel programming model, and automatically collects details about available computing resources using tools like hwloc.

Compeanu et al. \cite{compeanu2017} introduce a component-based approach for parallel computing software development on CPU-GPU embedded systems. They propose using APIs to abstract the characteristics of CPUs and GPUs, making the components platform-agnostic. The decision of running components on CPUs or GPUs is deferred to a higher system level. In contrast, our approach focuses on reusing existing components (implementation variants) with minimal modifications. Our pre-compiler ensures compatibility with other compilers, avoiding functionality disruptions if the code is not processed through our framework.

Carvalho et al. \cite{Carvalho2007} introduce the \# component model, enabling developers to build software using components. These components utilize Haskell-like parallel programming extensions, allowing for the separation of concerns and emphasizing processing separation from the developer's perspective. The components are synthesized into the desired parallel program by a back-end system. While our paper also focuses on systems built from components, we differ in our objective. Instead of synthesizing a complex application from smaller components, we optimize systems that feature multiple implementations of the same components.

Mani and Kesselman \cite{mani2005} present the Compositional C++ extensions for constructing parallel systems using components. Their focus is on providing extensions and features that facilitate the development of components themselves. In contrast, our work assumes that the components already exist and concentrates on providing mechanisms to effectively expose the multiple implementation variants of these components to both the compiler and the runtime system.

\section{Conclusion and future work}
\label{sec:conclusion_fw}

In this paper, we presented COMPAR, a framework for component-based parallel programming on heterogeneous systems. COMPAR allows developers to annotate multiple implementation variants of components using compiler directives, enabling the runtime system to select the most suitable variant at runtime. Through evaluation experiments, we demonstrated that COMPAR, in combination with the StarPU runtime system, effectively selects the best implementation variant for different scenarios. Despite the need for further training of the performance models, COMPAR offers a flexible and efficient approach to optimize performance on heterogeneous systems while maintaining compatibility with existing codebases. Future work involves enhancing the training of performance models, exploring integration with other runtime systems, and expanding its applicability to different programming models and architectures.

 \bibliographystyle{splncs04}

\end{document}